\documentclass{hpc-ua}

\begin{document}

\usefont{T2A}{ftm}{m}{n}
\selectlanguage{english}

\newcommand\ttot{{T_{\rm TOT}}}
\newcommand\thost{{T_{\rm host}}}
\newcommand\tgpu{{T_{\rm GPU}}}
\newcommand\tcomm{{T_{\rm comm}}}
\newcommand\tmpi{{T_{\rm MPI}}}

\newcommand\TS{n_{\rm ts}}
\newcommand\Nact{N_{\rm act}}
\newcommand\Nactsum{\sum \Nact}
\newcommand\Nactavr{\langle \Nact \rangle}
\newcommand\Ngpu{N_{\rm GPU}}

\title{Up to 700k GPU cores, Kepler, and the Exascale future for simulations of 
star clusters around black holes}

\author{Peter Berczik\affiliationmark{1,2,3}, 
        Rainer Spurzem\affiliationmark{1,3}, \\
				Long Wang\affiliationmark{4}, 
				Shiyan Zhong\affiliationmark{1}, 
				Siyi Huang\affiliationmark{1}, 
				Maxwell Xu Tsai\affiliationmark{1}, \\
				Gareth Kennedy\affiliationmark{1}, 
				Shuo Li\affiliationmark{1}, 
				Luca Naso\affiliationmark{1}, 
				Changhua Li\affiliationmark{1}, \\
				Alexander Veles\affiliationmark{2}, 
				Igor Zinchenko\affiliationmark{2}
				}

\affiliation{%
\affiliationmark{1}National Astronomical Observatories, Chinese Academy of Sciences, 
20A Datun Rd., Chaoyang Distr., Beijing 100012, P.R. China \\
\affiliationmark{2}Main Astronomical Observatory, National Academy of Sciences of 
Ukraine, MAO/NASU, 27 Akademika Zabolotnoho St. 03680 Kyiv, Ukraine \\
\affiliationmark{3}Zentrum fur Astronomie, Astronomisches Rechen-Institut, Univ. 
of Heidelberg, Monchhofstr. 12-14, 69120 Heidelberg, Germany \\
\affiliationmark{4}Kavli Inst. for Astronomy and Astrophysics, Peking University, 
Yi He Yuan Rd. 5, Haidian Distr., Beijing 100871, P.R. China
}

\email{berczik@mao.kiev.ua, berczik@nao.cas.cn, spurzem@nao.cas.cn}

\maketitle

\begin{abstract}
We present direct astrophysical N-body simulations with up 
to a few million bodies using our parallel MPI/CUDA code on large 
GPU clusters in China, Ukraine and Germany, with different kinds 
of GPU hardware. These clusters are directly linked under the 
Chinese Academy of Sciences special GPU cluster program in the 
cooperation of ICCS (International Center for Computational Science).
We reach about the half the peak Kepler K20 GPU performance for our 
$\varphi$-GPU\footnote{\tt ftp://ftp.mao.kiev.ua/pub/berczik/phi-GPU/} 
code ~\cite{BNZ2011}, in a real application scenario with individual 
hierarchically block time-steps with the high (4$^{th}$, 6$^{th}$ and 
8$^{th}$) order Hermite integration schemes and a real core-halo density 
structure of the modeled stellar systems. The code and hardware 
are mainly used to simulate star clusters ~\cite{KBP2009, JBP2009} and 
galactic nuclei with supermassive black holes ~\cite{JYM2012}, in which 
correlations between distant particles cannot be neglected.
\end{abstract}


\section{Introduction}


\begin{figure}[hbp]
\begin{center}
\includegraphics[angle=0,width=0.84\textwidth]{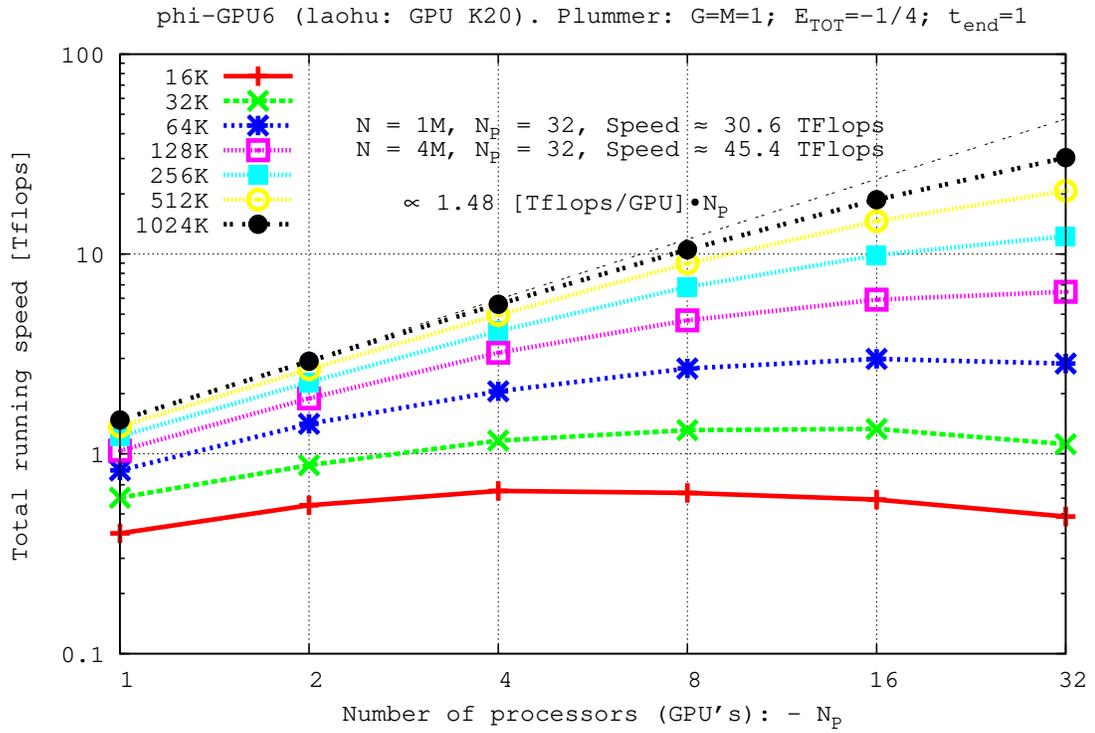}

\caption{Speed performance with mixed (fp32 + fp64) precision of the 
$\varphi$-GPU 6$^{th}$ order scheme on the K20 GPU cards. The lines 
with different symbols presents the different particle numbers.}

\label{fig:gflops}
\end{center}
\end{figure}

\begin{figure}[hbp]
\begin{center}
\includegraphics[angle=0,width=0.84\textwidth]{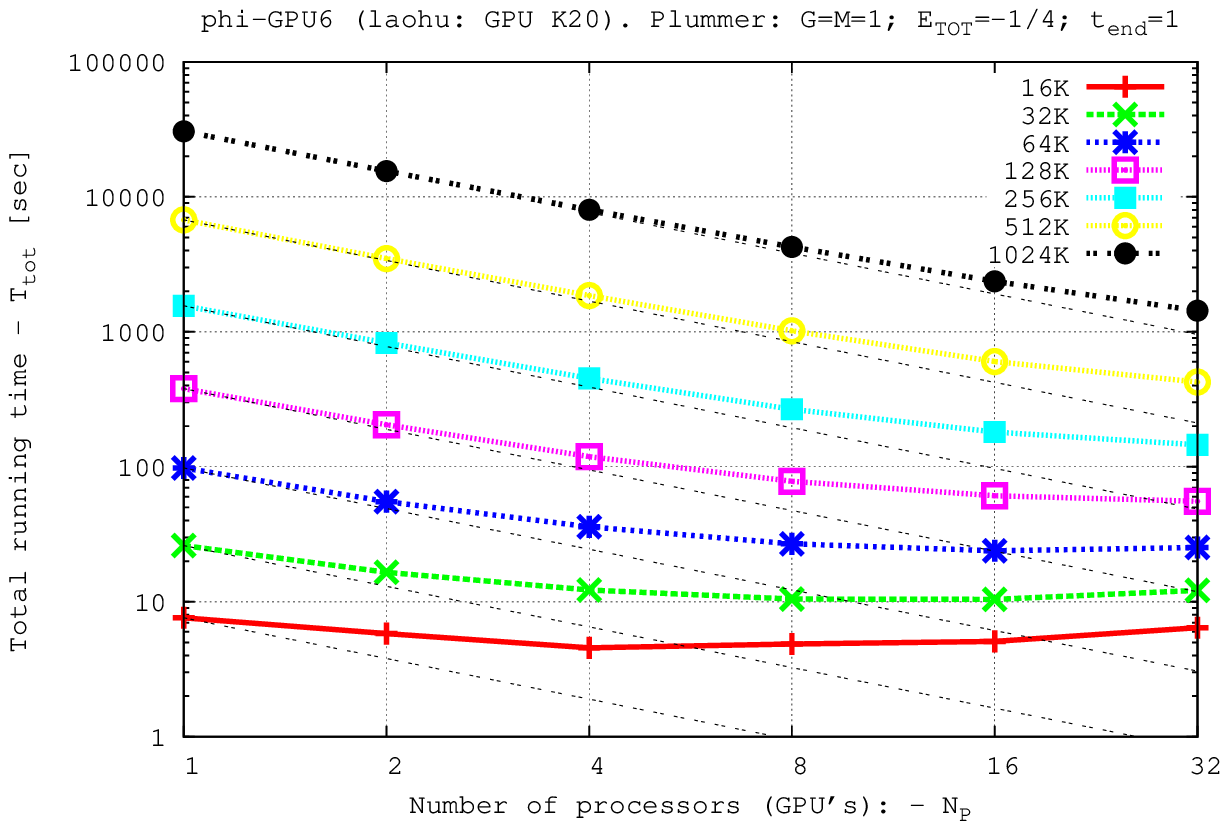}

\caption{Total wall clock time of 1 time unit integration with the 
$\varphi$-GPU 6$^{th}$ order scheme on the K20 GPU cards. The lines 
with different symbols presents the different particle numbers.}

\label{fig:time}
\end{center}
\end{figure}

Many, if not all galaxies harbor supermassive black holes 
~\cite{FPB2012, MBL2007}. If galaxies merge, which is quite common in the 
process of hierarchical structure formation in the universe, their black 
holes sink to the center of the merger remnant and form a tight binary 
~\cite{LLB2012, JKB2011, BMS2006, BMS2005}. Depending on initial conditions 
and time supermassive black hole binaries are prominent gravitational wave 
sources, if they ultimately come close together and coalesce 
~\cite{KHBJ2013, KPB2012, PBB2011}. We model such systems as gravitating N-body 
systems (stars) with two or more massive bodies (black holes), including if 
necessary relativistic corrections to the classical Newtonian gravitational 
forces, and model their gravitational radiation emission directly from the 
simulation ~\cite{KBB2012, BPB2009, PBB2009, BPB2008}.

Competitive astronomical and astrophysical research requires access 
to competitive computing facilities. Theoretical numerical modeling of
astrophysical objects, their composition, radiation, and dynamical
evolution has become a third basic method of astrophysical research,
besides observation and pure theory ~\cite{EJB2011}. Numerical 
modeling allows one to compare theory with observational data in 
unprecedented detail, and it also provides theoretical insight into 
physical processes at work in complex systems ~\cite{EJB2010}. 
Similarly, data processing of astrophysical observations comprises 
the use of complex software pipeline to bring raw data into a form 
digestible for observational astronomers and ready for exchange and 
publication; these are, e.g., mathematical transformations like 
Fourier analysis of time series or spatial structures, complex 
template analysis or huge matrix-vector operations. Here fast 
access to and transmission of data, require supercomputing capacities. 

We are undergoing a new revolution on parallel processor technologies,
especially with regard to the Graphic Processing Units. GPU's have become 
widely used nowadays to accelerate a broad range of applications, including
computational physics and astrophysics, image/video processing,
engineering simulations, quantum chemistry\footnote{\tt http://gpgpu.org/}. 
Graphics processing units (GPU's) are rapidly emerging as a
powerful and cost-effective platform for high performance parallel
computing. Recent GPU's, such as the NVIDIA Fermi C2050 and Kepler 
K20 Computing Processors offer correspondingly 448 and 2496 processor cores 
with the extremely fast on-chip-memory, as compared to only 6 or 8 cores 
on a standard Intel or AMD CPU. In this paper we present the set of large 
scale N-body benchmarks of our astrophysical applications on a large Kepler 
and Fermi based GPU clusters.


\section{Codes description}

The application codes which we use for benchmarking here are direct N-body 
simulation codes for astrophysics, both using a high order Hermite integration 
scheme and hierarchical block time steps. One is called $\varphi$-GPU 
\cite{BNZ2011, SBZ2012}, it has been developed from our earlier published versions 
$\varphi$-GRAPE\footnote{\tt ftp://ftp.mao.kiev.ua/pub/berczik/phi-GRAPE/} 
(which used GRAPE hardware as a accelerator \cite{HGM2007}), the other is a 
legacy code named NBODY6++GPU \cite{A1999, S1999}, which has only 
recently been accelerated by GPU. Both use MPI (Message Passing Interface), 
and on each node use many cores of the special GPU hardware to compute 
gravitational forces between particles. Parts of the codes were developed in 
cooperation with Keigo Nitadori (RIKEN Japan) and Tsuyoshi Hamada (Nagasaki Univ. 
Japan). $\varphi$-GPU \cite{BNZ2011, SBZ2012} is written in C++ with MPI and CUDA-C. 
NBODY6++GPU is written in Fortran with CUDA-C extensions \cite{NA2012}. We 
present benchmarks on different GPU accelerated hardware, including Fermi and 
Kepler architectures. 


\section{Large scale astrophysical results}

\begin{figure}[hbp]
\begin{center}
\includegraphics[angle=0,width=0.84\textwidth]{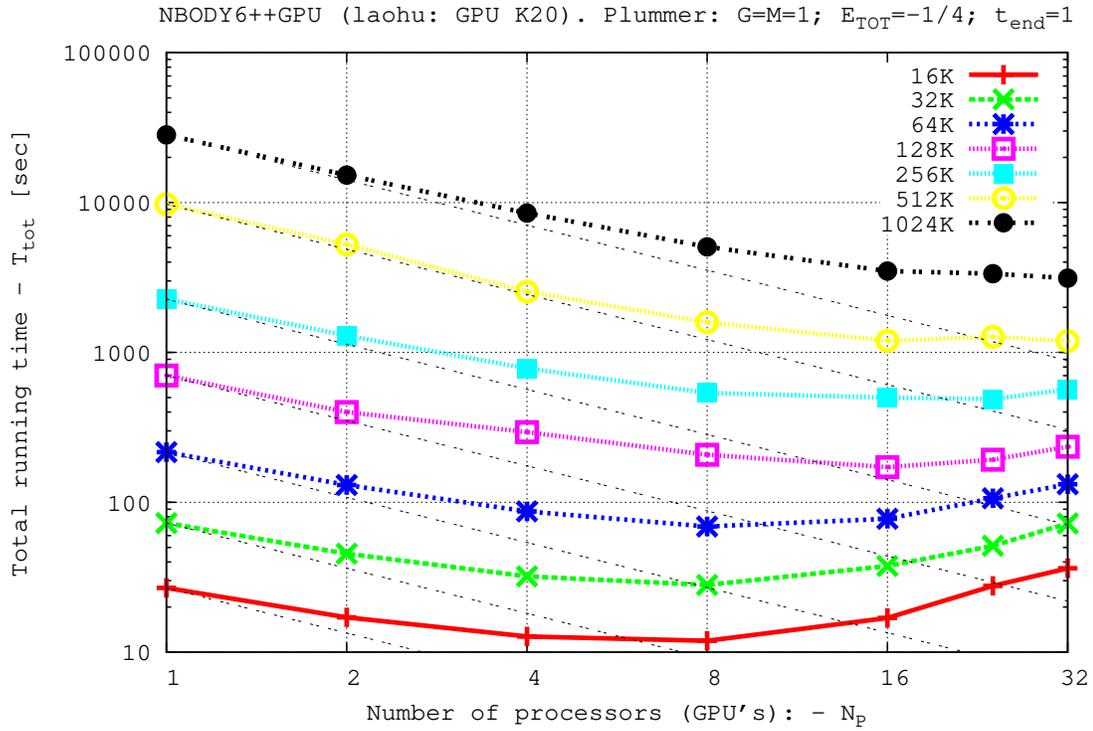}

\caption{Total wall clock time of 1 time unit integration with the 
NBODY6++GPU on the K20 GPU cards. The lines with different symbols 
presents the different particle numbers.}

\label{fig:nb6time}
\end{center}
\end{figure}

\begin{figure}[hbp]
\begin{center}
\includegraphics[angle=0,width=0.76\textwidth]{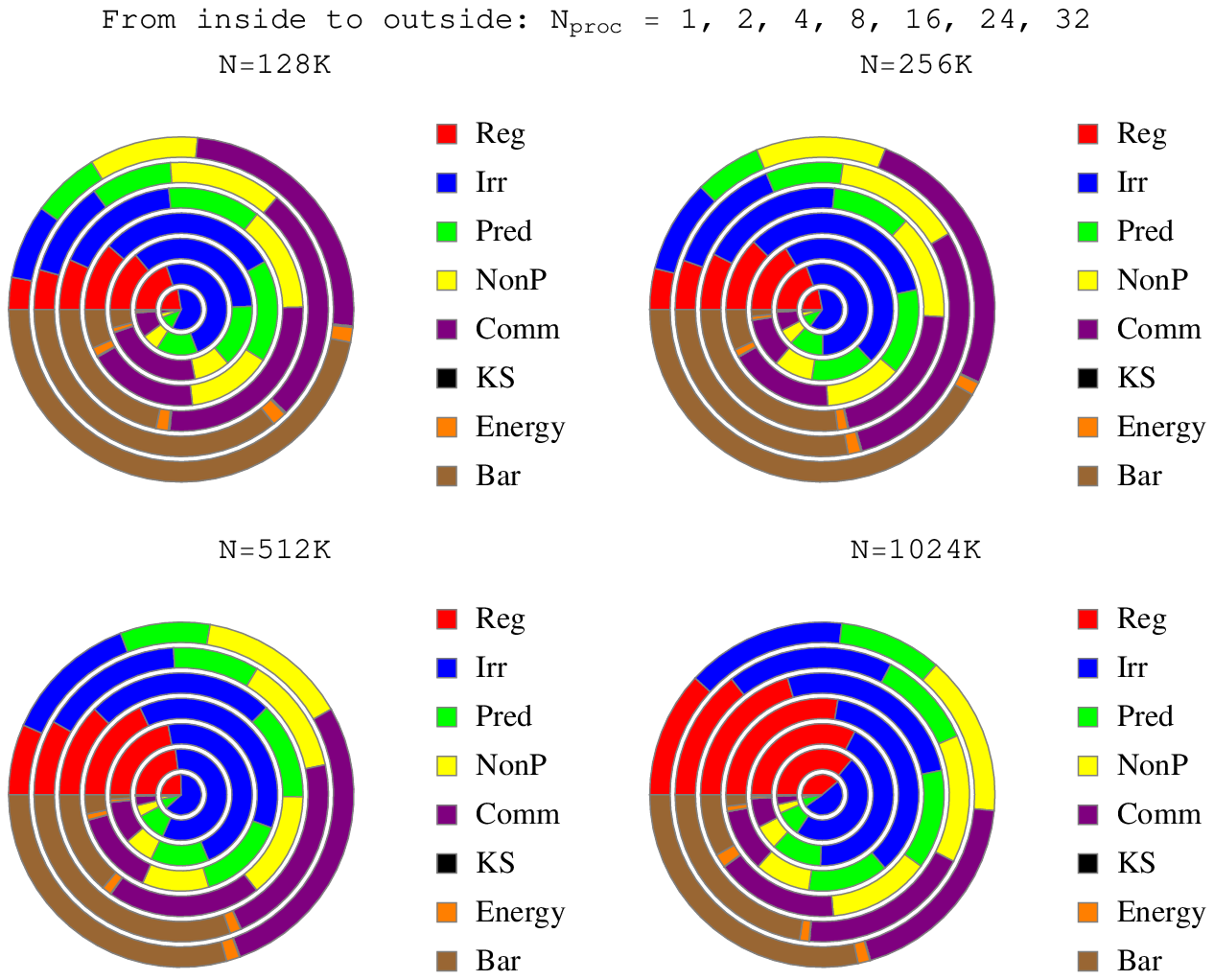}

\caption{Distribution of the total wall clock time of 1 time unit integration 
with the NBODY6++GPU on the K20 GPU cards. The different pie charts presents 
the code timing data for the different particle numbers.}

\label{fig:nb6pie}
\end{center}
\end{figure}

We report results obtained with our parallel direct N-body code on the CAS GPU 
clusters at our institute NAOC/CAS and at the IPE/CAS. We are also planning to 
run benchmarks on the Oak Ridge Titan supercomputer (using its Kepler K20 GPU's), 
but at the time of writing we cannot foresee whether they will be ready for the conference. 

\begin{figure}[hbp]
\begin{center}
\includegraphics[angle=0,width=0.84\textwidth]{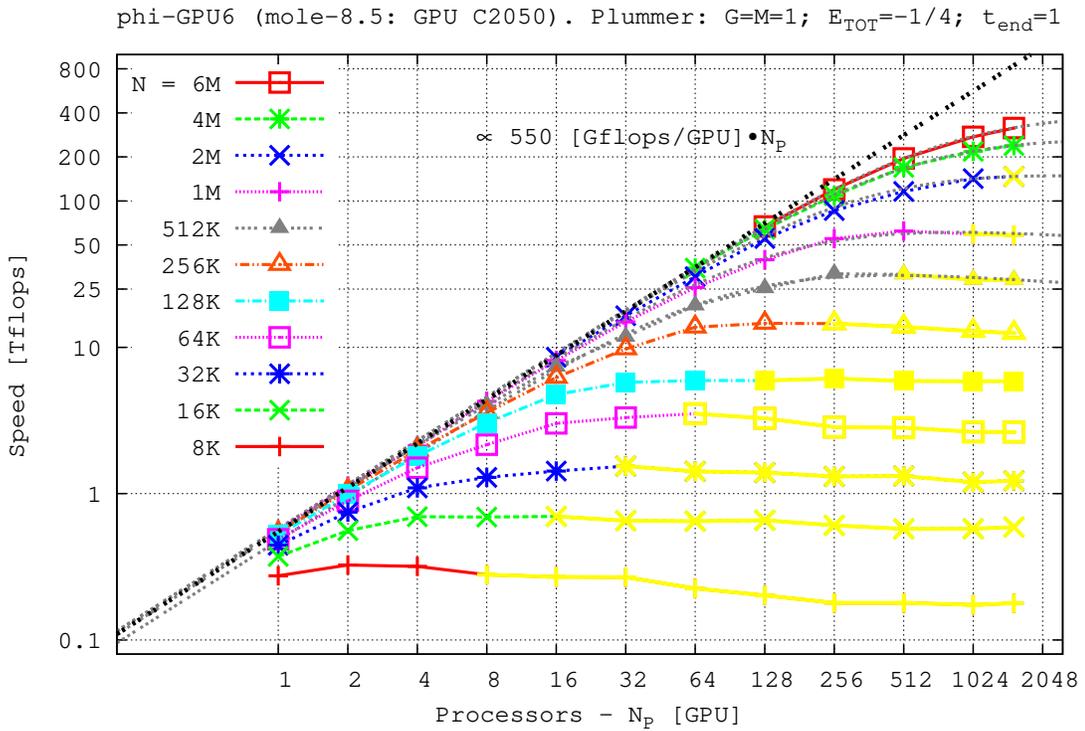}

\caption{Speed performance with mixed (fp32 + fp64) precision of the 
$\varphi$-GPU 6$^{th}$ order scheme on the C2050 GPU cards. The lines with 
different symbols presents the different particle numbers.}

\label{fig:res6}
\end{center}
\end{figure}

\begin{figure}[hbp]
\begin{center}
\includegraphics[angle=0,width=0.84\textwidth]{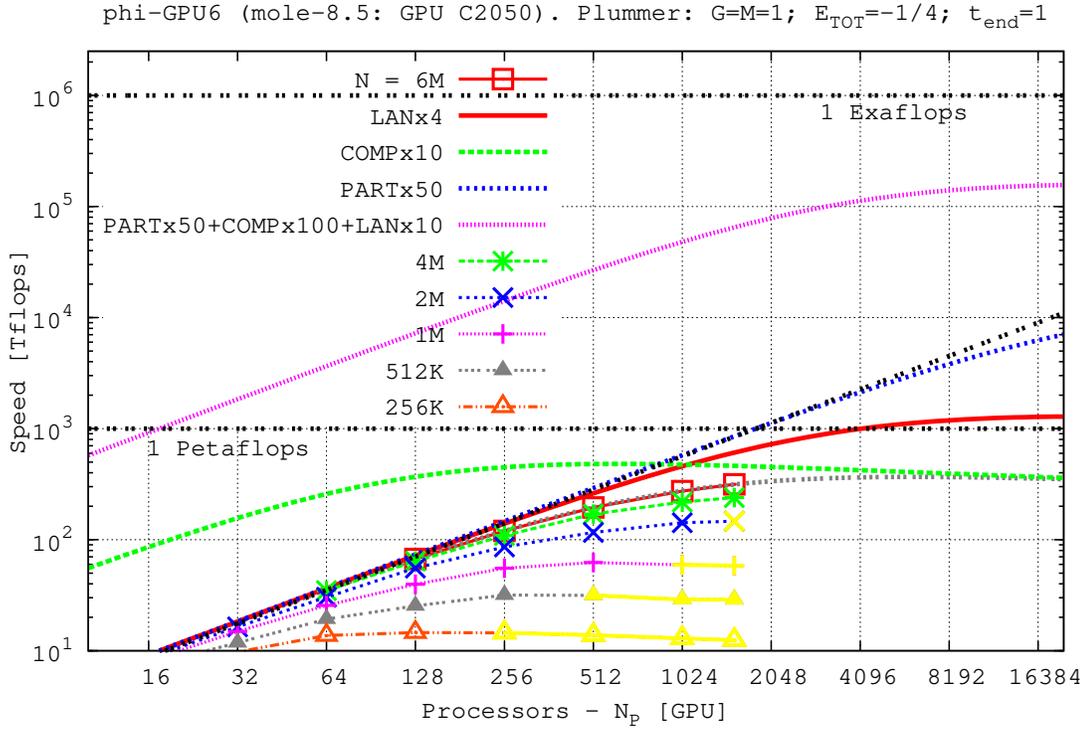}

\caption{Speed performance prognosis with mixed (fp32 + fp64) precision of 
the $\varphi$-GPU 6$^{th}$ order scheme on the C2050 GPU cards. The lines with 
different symbols presents the different particle numbers. The dotted horizontal 
lines shows the 1 Petaflops \& 1 Exaflops levels.}

\label{fig:res6-prog}
\end{center}
\end{figure}

The NAOC cluster (85 nodes) {\tt laohu} is now, after a recent upgrade, running 
with the new Kepler K20 GPU's (each with 2496 cores), and has in total roughly 
$\sim$170k computing GPU cores. The IPE/CAS system {\tt mole-8.5} is a much larger system 
with 362 nodes (each with 6 Fermi C2050 accelerators) and has in total almost $\sim$1M (973k) 
GPU cores. On the newest Kepler K20 GPU's we reach almost 1.5 Tflop/s sustained speed 
per one GPU using the 6$^{th}$ order $\varphi$-GPU code (see {\bf Figure~\ref{fig:gflops}.} 
and {\bf Figure~\ref{fig:time}.}). On the Fermi C2050 cards on the {\tt mole-8.5} 
system we get the sustained speed of around 550 Gflop/s per one GPU with the same 
$\varphi$-GPU code (see {\bf Figure~\ref{fig:res6}.} and {\bf Figure~\ref{fig:res6-prog}.}). 
These numbers are roughly close to the 50\% of the peak performance of such a GPU cards, 
which shows a very good utilization of the GPU hardware with our high order direct N-body 
codes. Note that for the largest simulation conducted for this system used 1536 Fermi C2050 
accelerators each with 448 computing cores totaling almost 700k GPU cores of the {\tt mole-8.5} system. 

After our extensive scaling measurement on the {\tt mole-8.5} system, which we have presented 
in our earlier paper \cite{BNZ2011}, we make a detail performance analysis of our $\varphi$-GPU 
code and derive the ``semi-analytic'' scaling formula for the performance (see also the 
equations (17), (18) and (19) in the paper \cite{BNZ2011}):

\begin{equation}
P \approx \frac{\gamma~N^{2+x}}
{\frac{\alpha N^{2+x}}{\Ngpu} + \beta (N^{0.33+x} \tau_{\rm lat} + N^{1+x}) \log(\Ngpu) } 
\label{eq:def_P3}
\end{equation}

The formula (based on the experimental data fitting) quite well predict the code speed for 
different combination of particle and processor numbers (see the {\bf Figure 4.} in the 
paper \cite{BNZ2011}). Based on this and using the extrapolation we predict our $\varphi$-GPU 
code performance for the much larger system (like the Oak Ridge Titan supercomputer system 
with $\sim$20k GPU accelerator cards). 

In the {\bf Figure~\ref{fig:res6-prog}.} we present this extrapolation exercise. We first 
simply extend the scale of the N=6M particle simulation for the larger GPU accelerator 
numbers up to 20k. As we can see even for this very high particle number the maximum 
performance which we can reach with the current GPU and network hardware is around 
$\sim$350 - 400 Tflop/s (see the line with N=6M label). We can have a significantly better 
speedup factor if we were to use the roughly 4 times faster network (see the line with LANx4 
label). In this case we can reach above the 1 Pflop/s speed. If we could simultaneously 
speedup the network 10 times and also speedup the GPU hardware with 100 times we can 
reach sub Exaflop/s speed for N=300M particles. Such a very large particle number 
is needed to describe the full galactic bulge system, consisting of a few 10$^8$ M$_\odot$, 
in star by star way. 

The second set of the tests were done on the {\tt laohu} cluster using the newly 
developed GPU enabled NBODY6++GPU code. This code is more complex including many binaries, 
stellar evolution and an Ahmad-Cohen neighbor scheme, and will be used for high resolution 
star cluster modeling. The results are shown in the plots 
{\bf Figure~\ref{fig:nb6time}.} and {\bf Figure~\ref{fig:nb6pie}.}. 
The new NBODY6++GPU code also shows a very good GPU utilization performance data 
and quite good network communication scaling. Right now, we are still working on 
the NBODY6++GPU code optimization, so, these results should be considered as 
``preliminary''. After the next round of network communication optimization, we hope to 
get a speedup factor of $\sim$2 especially for larger particle and processor numbers. 

As examples of the science applications done with these codes we will present 
new simulations for binary black holes to reach relativistic coalescence after 
galaxy mergers \cite{KHBJ2013, KBB2012, KPB2012}. The simulations show unexpectedly 
short times to reach full coalescence of the black holes, compared to the 
typical galaxy merging time (less than a Gyr). In another paper we show how the dynamics 
of binary black holes before the final merger gives rise to significant 
gravitational wave emission in the pulsar timing band, due to their high 
eccentricities \cite{PBB2011, BPB2009}. 


\section{Conclusions}

Our massively parallel codes ($\varphi$-GPU and NBODY6++GPU), which use MPI 
parallelization as well as acceleration by many GPU's, scale well on large 
numbers of cores. They both run very well with no sign of saturation e.g. by 
communication on the new Kepler K20 GPU accelerator, reaching almost 
1.5 Tflop/s per GPU with 2496 cores. These codes are currently being used for 
astrophysical research on galactic nuclei, requiring large particle resolution. 
With realistic technical improvements of GPU hardware and network speed we 
expect to reach approximately 0.2 Exaflop/s speed for N=300M particles.


\section{Acknowledgments}

We acknowledge support by Chinese Academy of Sciences through the Silk Road 
Project at NAOC, through the Chinese Academy of Sciences Visiting Professorship 
for Senior International Scientists, Grant Number $2009S1-5$ (RS), and through 
the "Qianren" special foreign experts program of China. 

The special GPU accelerated supercomputer {\tt laohu} at the Center of 
Information and Computing at National Astronomical Observatories, Chinese 
Academy of Sciences, funded by Ministry of Finance of People's Republic 
of China under the grant $ZDYZ2008-2$. We also used smaller GPU clusters  
{\tt titan}, {\tt hydra} and {\tt kepler}, funded under the grants 
I/80041-043 and I/84678/84680 of the Volkswagen Foundation and grants 
823.219-439/30 and /36 of the Ministry of Science, Research and the Arts 
of Baden-W\"urttemberg, Germany. Some code development was also done on 
the Milky Way supercomputer, funded by the Deutsche Forschungsgemeinschaft 
(DFG) through Collaborative Research Center (SFB 881) "The Milky Way System" 
(subproject Z2), hosted and co-funded by the J\"ulich Supercomputing Center (JSC). 

PB acknowledges the special support by the NASU under the Main Astronomical
Observatory GRID/GPU computing cluster project. 


\end{document}